\documentclass[
prper,
aps,
 twocolumn,
groupedaddress,floats,showpacs,final,superscriptaddress]{revtex4-2}
\usepackage{graphicx}
\usepackage{dcolumn}
\usepackage{bm}
\usepackage{amssymb}
\usepackage{epstopdf}
\usepackage{color}
\usepackage{amsmath}
\usepackage{ulem}

\begin{document}


\title{Electron diffusion induced valley Hall effect and nonlinear galvanodiffusive transport in hexagonal 2D Dirac monolayer materials}

\author{A.~V.~Snegirev}
\affiliation{Rzhanov Institute of Semiconductor Physics, \\
Siberian Branch of the Russian Academy of Sciences, Novosibirsk 630090, Russia}
\affiliation{Novosibirsk State University, Novosibirsk 630090, Russia}

\author{V.~M.~Kovalev}
\affiliation{Rzhanov Institute of Semiconductor Physics, \\
Siberian Branch of the Russian Academy of Sciences, Novosibirsk 630090, Russia}
\affiliation{Novosibirsk State Technical University, Novosibirsk 630073, Russia}

\author{M.~V.~Entin}
\affiliation{Rzhanov Institute of Semiconductor Physics, \\
Siberian Branch of the Russian Academy of Sciences, Novosibirsk 630090, Russia}

\begin{abstract}
Diffusion currents are theoretically examined in two-dimensional Dirac materials, such as those of the transition metal dichalcogenides (TMD) family. The transversal effects are analogues of the valley Hall (VHE) and photogalvanic (PGE) transport phenomena in case when the electron driving force is not an electric field but a gradient of electron density distribution in the sample. The latter can be created by a finite-sized laser spot or by the injection of electrons from other materials. We develop the theory of diffusive VHE effect assuming the anisotropic electron-short-range-impurity skew scattering. The electron PGE-like transport caused by higher electron-density derivatives is analyzed assuming the trigonal warping anisotropy of electron valleys in a TMD monolayer. The nonlinear responses on electron-density gradient are studied as well. The isotropic processes of electron scattering off the short-range and Coulomb centers are taken into account in the PGE-like transport theory.
\end{abstract}

\maketitle

\section{Introduction}
Hexagonal two-dimensional monolayer structures, in particular, graphene and transition metal dicholcogenides (TMD) \cite{Mak}, are under active study nowadays. The research of their physical properties occurs in two main directions: optical properties \cite{glazovReview, Wang} and transport phenomena \cite{transport1,transport2,transport3,transport4}.  Exceptional optical properties of TMD based on a specific structure of excitons and exciton-polaritons in TMDs have wide perspectives in optoelectronics \cite{glazovReview}.

The transport properties of monolayer TMD materials also demonstrate specific features due to their unique band structure \cite{band, band2}. TMD monolayers have the hexagonal lattice structure belonging to the D$_{3h}$ point group producing the two-valley structure of their Brillouin zone having a strong valence band spin-orbit splitting. The specific optical interband selection rules, allowing for the driving of the valley degrees of freedom of the charge carriers producing very specific transport properties of these materials and the corresponding research direction called \textit{valleytronics} \cite{valleytronics} in modern literature have recently arisen.

The transport response of TMD materials under study can be split phenomenologically by the amplitude ${\bf E}$ of external electromagnetic wave into a linear and second order current density response. The linear transport phenomena, being first order with respect to ${\bf E}$, obey the relation $j_\alpha=\sigma_{\alpha\beta}E_\beta$, where $j_\alpha$ are current density components and $\sigma_{\alpha\beta}$ are kinetic coefficients being the components of generalized  conductivity tensor. The most intriguing linear transport phenomena in TMDs is the existence of valley Hall effect (VHE) \cite{VHE1, VHE1.1, VHE2, VHE3}: the transverse current density rose as a linear response to the in-plane static electric field ${\bf E}$ and was determined by the non-diagonal components of conductivity tensor as $\sigma_{H}=\sigma_{xy}=\sigma_{yx}$. The specific feature of VHE is its valley-selective nature: the transverse VHE current densities flow in opposite directions in different valleys in such a way that the net transverse current, being summed over two valleys, vanishes. This is due to the time-reversal symmetry by which the valleys are coupled. The net VHE current can be viewed when the time-reversal symmetry is destroyed. The latter can be done either by the sample illumination by the circularly polarized electromagnetic field or by the external magnetic field. The experimental observation of VHE in a TMD-based transistor structure was made by the first method \cite{VHE2}. The microscopic mechanisms producing the VHE transport were intensively studied theoretically. Initially, VHE was predicted in \cite{VHE1}, where the Berry-phase-induced anomalous contribution to the electron velocity was considered as the underlying microscopic mechanism of VHE \cite{VHE1.1}. Later on, the key role of anomalous electron scattering processes off the impurity potential in the theoretical understanding of VHE was recognized \cite{VHE4, VHE5, VHE6}.

Recently, a detailed theoretical description for VHE was developed in \cite{VHEGlazovGolub, VHEcleanGlazov} considering also the skew and side-jump electron scattering off the impurities in dirty samples. The authors considered also different electrons driving mechanisms, the standard one, due to the in-plane static electric field ${\bf E}$, and also due to the photon \cite{VAEE} and phonon electron dragging. It was shown that the dominant microscopic mechanism of VHE in disordered samples is determined by the skew-electron scattering off impurities. Later, the theoretical description of non-equilibrium VHE, including both impurity scattering and the interband valley-selective illumination, was developed \cite{noneqVHE}. The regime of strong interband transitions, in the case of the Berry-phase-induced anomalous velocity VHE mechanism, was also analyzed \cite{StrongBerryVHE}. It was shown that, in these cases, the skew-related electron scattering gives the dominant contribution to VHE, which made it possible to explain some observations of the VHE transport experiment \cite{VHE2}.

The other transport phenomenon actively studied in TMD materials is the nonlinear effect, say, photogalvanic effect (PGE) \cite{PGE1, PGE2, PGE3, PGE4}. The PGE is the appearance of stationary (and also uniform in space) current density due to the external alternating electromagnetic field with a normal incidence to the TMD plane and, phenomenologically, can be written as $j_{\alpha}=\chi_{\alpha\beta\gamma} E_\beta E^*_\gamma$. The PGE effect is well known in 3D semiconductor physics \cite{PGE1, PGE2} and it was well understood in conventional semiconductors \cite{deyo, golubivchenkospivak, PGE4}. In TMD materials, PGE may arise both due to the trigonal symmetry of electron valleys or also due to anomalous velocity (Berry-phase) and electron-impurity scattering. The PGE in TMD monolayers also has the valley-selective nature: the PGE current flow in opposite directions in different valleys \cite{PGEEntin1, PGEEntin2}. Destroying the time-reversal symmetry produces the nonzero net PGE current in the sample \cite{PGEEntin1, PGEEntin2}.

In all transport effects discussed above, the driving force was the electric field: it is stationary in the plane field or an alternative electric field of electromagnetic wave. From the general point of view \cite{blokhentin}, any factor, which drives the system from equilibrium state, may produce the current. Any such a factor, called generalized force in statistical physics, may be a scalar, vector or even a tensor. As a vector, a generalized force may be an electric field, a temperature gradient and particle density gradients, whereas a scalar generalized force may be a temperature difference \cite{Tarasenko} or particle concentration difference between different subsystems or their time derivatives. The tensor generalized force may be constructed from the second spatial derivatives of the scalar quantities etc.  Besides, the current can be excited by the higher orders of the vector forces together with their cross products.  A general expression for  currents caused by scalar $F$,  vector $F_j$, and tensor $F_{ij}$ forces is
\begin{gather}\label{Hall00}\nonumber
j_i=\gamma^{(0)}_iF+\gamma^{(1)}_{ij}\partial_j F+ \gamma^{(2)}_{ij} F_j+\gamma^{(3)}_{ijk} F_{j} F_{k}+\gamma^{(4)}_{ijk} F_{jk}+…
\end{gather}
Besides, the cross terms due to different forces are possible in higher orders of generalized forces. The symmetry of the system restricts the possibility of coefficients $\gamma^{(0)}_i$, $\gamma^{(1)}_{ij}$, $\gamma^{(3)}_{ijk}$, $\gamma^{(4)}_{ijk}$. The quantity $\gamma^{(0)}_i$ exists in pyroelectrics, the tensors $\gamma^{(3)}_{ijk}\neq0$ or $\gamma^{(4)}_{ijk}\neq 0$ demand the absence of reflection symmetry.

In case of vector generalized force, the general expression for the current density may be written as
\begin{gather}\label{Hall0}\nonumber
j_i=\alpha_{ij}F_j+\beta_{ijk}F_jF_k+... .
\end{gather}
The aim of the present paper is the theoretical description of VHE and PGE transport phenomena when electrons drive a vector generalized force given by the nonuniform electron density distribution in the sample, the particle density gradient, ${\bf \nabla}\cdot n({\bf r})$.  The latter can be created by the sample illumination with a finite-in-plane laser spot or by an injection of electrons from other materials.  From the phenomenological point of view, the diffusive-induced VHE reads as $j_{\alpha}({\bf r})=-eD_{\alpha\beta}\nabla_{\beta}n({\bf r})$ with generalized diffusive coefficients, $D_{\alpha\beta}$ ($e$ is an electron charge). The VHE will be given by their non-diagonal elements. The PGE-like current can be written as a nonlinear response to the density gradient, $j_{\alpha}({\bf r})=\chi_{\alpha\beta\gamma}\nabla_{\beta}\nabla_{\gamma}n({\bf r})+\zeta_{\alpha\beta\gamma}\nabla_{\beta}n({\bf r})\nabla_{\gamma}n({\bf r})$.

The paper has the following structure. In the next section we derive the non-diagonal elements of $D_{\alpha\beta}$, considering the electron-impurity skew scattering as the dominating mechanism. The impurities are considered to be of the short-range type. The later sections are devoted to the derivation of $\chi_{\alpha\beta\gamma}$ and $\zeta_{\alpha\beta\gamma}$ tensors. As the microscopic mechanism of diffusive PGE-like transport, we consider the trigonal warping of the electron valley dispersion in TMDs; both short-range and Coulomb impurities are analyzed in that sections.

\section{Valley Hall effect due to electron diffusion}
Here we consider the VHE effect due to the electron diffusion based upon the Boltzmann transport equation approach. We assume that the nonuniform electron density, which can be excited by the external electromagnetic field (or injected to the sample), acquires the fast energy relaxation and the resulting electron density is characterized by the quasi-equilibrium Fermi distribution function with a chemical potential being the arbitrary function of coordinates. It reads $f({\bf r})=(1+\exp[\varepsilon({\bf p})-\mu({\bf r})]/T)^{-1}$, where $\varepsilon({\bf p})$ is an electron dispersion in the given valley and $\mu({\bf r})$ is an electron chemical potential. We assume that the spatial non-uniformity of the electron distribution function (and corresponding electron density $n({\bf r})$) is weak in comparison with uniform electron density $N$ in the sample, $|n({\bf r})-N|\ll N$. We assume also that uniform distribution function $f_0=(1+\exp[\varepsilon({\bf p})-\mu_0]/T)^{-1}$ corresponds to the uniform electron density value $N$ in the sample.

The diffusion-like VHE current arises due to the nonuniform electron distribution and is determined by the first order correction $f^{(1)}({\bf r})$ to the electron distribution function, with respect to gradients, satisfying the following Boltzmann equation
\begin{gather}\label{Hall1}
{\bf v}\frac{\partial f({\bf r})}{\partial {\bf r}}+Q\{f^{(1)}({\bf r})\}=0,
\end{gather}
where ${\bf v}$ is an electron velocity in a given valley, $Q$ is an electron-impurity collision integral. Further in this section we derive all expressions for one given valley. The electron dispersion is assumed in this section to have the parabolic form $\varepsilon({\bf p})=p^2/2m$, and ${\bf v}={\bf p}/m$, respectively. Collision integral $Q$ consists of two terms describing the isotropic $Q^s$ and anisotropic $Q^a$ electron-impurity scattering. The isotropic scattering is approximated by the relaxation-time approach $Q\{f^{(1)}({\bf r})\}=f^{(1)}({\bf r})/\tau$. For simplicity we set $\tau$ to be independent of the electron energy. The first order correction to the distribution function can be also split into symmetric and antisymmetric contributions, $f^{(1)}=f^{(1)}_s+f^{(1)}_a$, with respect to the electron momentum. The VHE current density is expressed via the antisymmetric part as
\begin{gather}\label{Hall2}
j_\alpha=e\int\frac{d^2{\bf p}}{(2\pi\hbar)^2}v_{\alpha}f^{(1)}_a.
\end{gather}
Assuming the anisotropic scattering to be weak, the antisymmetric contribution to the electron distribution function can be found by a successive approximation. Thus, we have the following set of equations
\begin{gather}\label{Hall3}
{\bf v}\frac{\partial f_0}{\partial {\bf r}}+\frac{f^{(1)}_s}{\tau}=0,
\end{gather}
for the symmetric part and
\begin{gather}\label{Hall4}
\frac{f^{(1)}_a}{\tau}+Q^a\{f^{(1)}_s\}=0,
\end{gather}
for the antisymmetric one. Solving these equations, we find 
\begin{gather}\label{Hall5}
j_\alpha({\bf r})=e\tau^2\int\frac{d^2{\bf p}}{(2\pi\hbar)^2}v_{\alpha}
Q^a\{v_\beta\nabla_\beta f({\bf r})\}.
\end{gather}
An asymmetric part of the collision integral responsible for the skew scattering, $Q^a$, for a TMD monolayer was found in \cite{VHEGlazovGolub} and reads
\begin{gather}\label{Hall6}
Q^a\{F({\bf p})\}=W_0\sum_{{\bf p}'}F({\bf p}')[{\bf p}\times{\bf p}']_z\delta(\varepsilon_{\bf p}-\varepsilon_{{\bf p}'}),
\end{gather}
where $F({\bf p})$ is an arbitrary function of electron momentum. Parameter $W_0=2\pi u_0v^2/\tau\Delta^2$ is expressed via the short-range impurity potential amplitude modelled by a Dirac delta function as $U({\bf r})=u_0\delta({\bf r})$, $v$ is the TMD monolayer band parameter having the velocity dimensionality, $\Delta$ is the TMD bandgap and $\tau$ is the electron momentum relaxation time due to the electron-impurity scattering. A direct analysis based upon Eqs.\eqref{Hall5} and \eqref{Hall6} yields
\begin{gather}\label{Hall7}
j_y=2e\tau^2W_0\rho N\langle\varepsilon\rangle\left(-\frac{\partial \mu}{\partial x}\right),
\end{gather}
where $\rho=m/2\pi\hbar^2$ is an electron density of states, $N$ is an equilibrium electron density and
\begin{gather}\label{Hall8}
\langle\varepsilon\rangle=\frac{1}{N}\int\limits_0^\infty d\varepsilon\rho \varepsilon f_0
\end{gather}
is an average electron energy. Taking into account the relation between density and chemical potential gradients
\begin{gather}\label{Hall9}
\nabla n({\bf r})=\nabla \mu({\bf r})\int\frac{d^2{\bf p}}{(2\pi\hbar)^2}\left(-f_0'\right)=\\\nonumber
=\rho\left(1-e^{-\frac{N}{\rho T}}\right)\nabla \mu({\bf r}),
\end{gather}
one finds a VHE diffusive coefficient
\begin{gather}\label{Hall10}
D_{yx}=\frac{2\tau^2W_0N\langle\varepsilon\rangle}{1-e^{-\frac{N}{\rho T}}}=\frac{4\pi v^2\tau}{1-e^{-\frac{N}{\rho T}}}\frac{Nu_0\langle\varepsilon\rangle}{\Delta^2}.
\end{gather}
In Eq.\eqref{Hall9} we used the following approximation $\partial_\mu f({\bf r})=-f'({\bf r})\approx -f'_0$, where prime means derivative, with respect to the electron energy. A found expression Eq.\eqref{Hall10} holds for a given valley; in the other valley it has opposite sign.

\section{General expressions for PGE-like diffusion currents}

PGE-like current density, nonlinear with respect to electron density gradients, is phenomenologically expressed as
\begin{gather}\label{PGE1}
j_{\alpha}({\bf r})=\chi_{\alpha\beta\gamma}\nabla_{\beta}\nabla_{\gamma}n({\bf r})+
\zeta_{\alpha\beta\gamma}\nabla_{\beta}n({\bf r})\nabla_{\gamma}n({\bf r}).
\end{gather}
One can see, the current density is determined by the third-order tensors and, thus, it may occur in noncentrosymmetric systems. The TMD monolayer structures are described by the D$_{3h}$ point group which is of noncentrosymmetric class. In the D$_{3h}$ point group the non-zero elements of any third-order tensor read
\begin{gather}\label{PGE2}
-\chi_{xxx}=\chi_{xyy}=\chi_{yxy}=\chi_{yyx},
\end{gather}
whereas other component are zero. The same relation holds for the $\zeta_{\alpha\beta\gamma}$ tensor. Thus, it is enough to consider $\chi_{xxx}$ and $\zeta_{xxx}$ components only. We are interested in the PGE effect produced by the warping of the electron spectrum in TMD monolayers. Within the two-band model of electron dispersion of a TMD monolayer, the bare Hamiltonian accounting for the trigonal warping of electron valleys reads
\begin{gather}\label{PGE3}
H_0=\frac{\Delta}{2}\sigma_z+v(\eta\sigma_xp_x+\sigma_yp_y)+\left( \begin{array}{cc}0 & \mu p_+^2 \\ \mu p_-^2 & 0 \\ \end{array}\right),
\end{gather}
where $\Delta$ is the TMD material bandgap, $v=p_{cv}/m_0$, $\mu$ is the warping strength band parameter, $p_{\pm}=p_x\pm i p_y$ is an electron momentum and $\eta-\pm 1$ is a valley index. In the effective mass approximation, the conduction band electron dispersion near $K$ and $-K$ points of the Brillouin zone can be approximated as
\begin{gather}\label{PGE4}
\varepsilon_\textbf{p} = \varepsilon^0_\textbf{p}+w_{\bf p},\\\nonumber
\varepsilon^0_\textbf{p}=\frac{{\bf p}^2}{2m},\,\,\,w_{\bf p}=\eta{w}(p_x^3-3p_xp_y^2),
\end{gather}
where $w_{\bf p}$ is a warping correction to the electron dispersion in the $\eta-$th valley, where its strength is $w\sim\mu$. The electron distribution function now has to be found up to the second order, with respect to electron density gradients. Thus, the simple analysis of Boltzmann equation results in the PGE-like current density expression 
\begin{gather}\label{PGE5}
j_\alpha=e\int\frac{d{\bf p}}{(2\pi\hbar)^2}v_{\alpha}\hat{Q}^{-1}({\bf v}\nabla)\hat{Q}^{-1}({\bf v}\nabla)f^w({\bf r}),
\end{gather}
where quasiequilibrium distribution function $f^w({\bf r})$ contains now the electron dispersion with a warping correction, Eq.\eqref{PGE4}. The analytical theory can be developed assuming the smallness of the warping term $w_{\bf p}$ in the electron dispersion of Eq.\eqref{PGE4}. Thus, we will find the current by Eq.\eqref{PGE5} in the first order with respect to $w_{\bf p}$. The structure of Eq.\eqref{PGE5} dictates that the warping correction may come from i) the electron dispersion in distribution function $f^w({\bf r})$, ii) electron velocity $v_\alpha$ and, finally, iii) from the structure of collision operator $\hat{Q}$ \cite{PGEEntin1}. The energy dispersion acquiring the warping correction in the first order is given by Eq.\eqref{PGE4}, the structure of the electron velocity with the first order warping correction can be easily found as
\begin{gather}\label{PGE6}
{\bf v}={\bf v}^0+\delta{\bf v},\\\nonumber
{\bf v}^0=\frac{\bf p}{m},\,\,\,\,\delta{\bf v}=3\eta w(p_x^2-p_y^2,-2p_xp_y),
\end{gather}
whereas the correction to the collision operator caused by the warping term requires the careful analysis made below.

Now consider the warping correction to the collision operator $\hat{Q}$. An electron-impurity collision operator acting to the arbitrary function $\chi_{\bf p}$ in the lowest Born approximation yields
\begin{gather}\label{Q1}
\hat{Q}\{\chi_{\bf p}\}=\frac{2\pi n_i}{\hbar}\sum_{{\bf p}'}|M_{{\bf pp}'}|^2\delta(\varepsilon_{\bf p}-\varepsilon_{{\bf p}'})(\chi_{\bf p}-\chi_{{\bf p}'}),
\end{gather}
where $n_i$ is a short-range impurities density and $M_{{\bf pp}'}$ is a scattering matrix element of impurity potential. Following \cite{PGEEntin1} we neglect here the warping-induced corrections to the scattering matrix elements, and thus, $\hat{Q}$ acquires the correction due to the presence of electron dispersion $\varepsilon_{\bf p}$ in the energy conservation law at the electron-impurity scattering. Thus, one finds $\hat{Q}_0+\hat{Q}_w$, where the action of bare collision operator $\hat{Q}_0$ on the $n-$th harmonic of the distribution function reads
\begin{gather}\label{Q2}
\lim_{\alpha\rightarrow0}\left(\alpha+\hat{Q}_0\right)^{-1}e^{in\varphi}=e^{in\varphi}\lim_{\alpha\rightarrow0}
\left(\alpha+1/\tau_n\right)^{-1},\\\nonumber
\frac{1}{\tau_n}=\frac{2\pi n_i}{\hbar}\sum_{{\bf p}'}|M_{{\bf pp}'}|^2\delta(\varepsilon^0_{\bf p}-\varepsilon^0_{{\bf p}'})(1-\cos n\theta),
\end{gather}
whereas the action of $\hat{Q}_w$ is determined as
\begin{gather}\label{Q3}
\hat{Q}_w\{\chi_{\bf p}\}
=\frac{2\pi n_i}{\hbar}\sum_{{\bf p}'}|M_{{\bf pp}'}|^2
(w_{\bf p}-w_{{\bf p}'})\times\\\nonumber\times
\delta'(\varepsilon^0_{\bf p}-\varepsilon^0_{{\bf p}'})
(\chi_{\bf p}-\chi_{{\bf p}'}).
\end{gather}
Here prime means derivative, with respect to delta-function argument. Now express the inverse collision operator up to the first order with respect to the warping term as
\begin{gather}\label{current1}
\left(\hat{Q}_0+\hat{Q}_w\right)^{-1}\approx
\hat{Q}_0^{-1}-\hat{Q}_0^{-1}\hat{Q}_w\hat{Q}_0^{-1},
\end{gather}
the $j_x$ PGE-like current density expression can be split into three contributions
\begin{gather}\label{current2}
j^I_x=e\int\frac{d{\bf p}}{(2\pi\hbar)^2}
v_{x}\hat{Q}^{-1}_0(v_x\partial_x)\hat{Q}^{-1}_0(v_x\partial_x)f^w({\bf r}),\\\nonumber
j^{II}_x=-e\int\frac{d{\bf p}}{(2\pi\hbar)^2}
v_{x}\hat{Q}_0^{-1}\hat{Q}_w\hat{Q}_0^{-1}(v_x\partial_x)\hat{Q}^{-1}_0(v_x\partial_x)f^w({\bf r}),\\\nonumber
j^{III}_x=-e\int\frac{d{\bf p}}{(2\pi\hbar)^2}
v_{x}\hat{Q}_0^{-1}(v_x\partial_x)\hat{Q}_0^{-1}\hat{Q}_w\hat{Q}_0^{-1}(v_x\partial_x)f^w({\bf r}).
\end{gather}
These expressions give the PGE-like current density. Below we analyse these expressions in case of electron scattering off the neutral short-range and charged Coulomb impurities, respectively.

\subsection{PGE-like current density: short-range impurities}
It is easy to show that if electrons scattered off the short-range impurity potential, the operator $\hat{Q}_w$ in Eq.\eqref{Q3} acting on the arbitrary function of electron momentum $\chi_{\bf p}$, gives zero
\begin{gather}\label{shortrange1}
\hat{Q}_w\{\chi_{\bf p}\}=0.
\end{gather}
Thus, both $j^{II}$ and $j^{III}$ current contributions in Eq.\eqref{current2} vanish. In the remaining term $j^{I}$, expanding the distribution function up to the first order, with respect to warping, is made with the following equation
\begin{gather}\label{shortrange2}
f^w({\bf r})=f({\bf r})-w_{\bf p}\partial_\mu f({\bf r}),
\end{gather}
where $f({\bf r})$ is a bare function with isotropic and parabolic $\varepsilon^0_{\bf p}$ electron dispersions. Keeping only $w_{\bf p}$-like first order terms, the PGE-like current density can be expressed via chemical potential gradients as
\begin{gather}\label{shortrange2.1}
j_{\alpha\beta\gamma}=A_{\alpha\beta\gamma}\nabla_\beta\nabla_\gamma\mu({\bf r})+B_{\alpha\beta\gamma}\nabla_\beta\mu({\bf r})\nabla_\gamma\mu({\bf r}),
\end{gather}
where
\begin{gather}\label{shortrange2.2}
A_{xxx} = e\eta{w}N
\Bigl[6\langle(\epsilon^2\tau_{1}\tau_{2})'\rangle
+3\langle(\epsilon^2\tau_{1}^2)'\rangle-
\langle(\epsilon^3\tau_{1}\tau_{2})''\rangle\Bigr],\\\nonumber
B_{xxx} = e\eta{w}N
\Bigl[6\langle(\epsilon^2\tau_{1}\tau_{2})''\rangle
+3\langle(\epsilon^2\tau_{1}^2)''\rangle-
\langle(\epsilon^3\tau_{1}\tau_{2})'''\rangle\Bigr],
\end{gather}
and $\tau_{n}$ is an electron relaxation time of $n-$th harmonics; prime here means a derivative with respect to electron bare energy $\varepsilon\equiv\varepsilon^0_{\bf p}$, and the spatially uniform energy distribution averaging is determined by Eq.\eqref{Hall8}.

The relation between tensors $A_{\alpha\beta\gamma}, B_{\alpha\beta\gamma}$ and $\chi_{\alpha\beta\gamma}, \zeta_{\alpha\beta\gamma}$ can be found in the general form for the arbitrary degeneracy of electron gas. It can easily be shown that
\begin{gather}\label{shortrange2.3}
\chi_{\alpha\beta\gamma}=\frac{A_{\alpha\beta\gamma}}{\sum_{\bf p}(-f_0')},\\\nonumber
\zeta_{\alpha\beta\gamma}=\frac{B_{\alpha\beta\gamma}\sum_{\bf p}(-f_0')-A_{\alpha\beta\gamma}\sum_{\bf p}f_0''}{\Bigl[\sum_{\bf p}(-f_0')\Bigr]^3},
\end{gather}
where
\begin{gather}\label{shortrange2.4}
\sum_{\bf p}(-f_0')=\rho\left(1-e^{-\frac{N}{\rho T}}\right),\\\nonumber
\sum_{\bf p}f_0''=\frac{\rho}{T}e^{-\frac{N}{\rho T}}\left(1-e^{-\frac{N}{\rho T}}\right).
\end{gather}
In systems where the electron dispersion is quadratic with respect to electron momentum, the relaxation time corresponding to the electron scattering off short-range impurities does not depend on the electron energy, $\tau_1=\tau_2=\tau=const$.  Thus, Eqs.\eqref{shortrange2.3} give
\begin{gather}\label{shortrange5}
\chi_{xxx} =\eta\frac{12ewN\tau^2}{\rho}\frac{\langle\epsilon\rangle}{1-e^{-\frac{N}{\rho T}}},\\\nonumber
\zeta_{xxx} = \eta\frac{12ewN\tau^2}{\rho^2}
\frac{1-\frac{\langle\epsilon\rangle}{T}e^{-\frac{N}{\rho T}}}{\left(1-e^{-\frac{N}{\rho T}}\right)^2}.
\end{gather}

\subsection{PGE-like current density: Coulomb impurities}
Consider the PGE-like current in the case of electron scattering off Coulomb impurities. Expression Eq.\eqref{Q1} is valid with the matrix element given by the Fourier-transformed 2D Coulomb potential
\begin{gather}\label{Сoulomb0}
M_{{\bf pp}'}=\frac{2\pi e^2\hbar}{\kappa |{\bf p}-{\bf p}'|},
\end{gather}
where $\kappa$ is the dielectric permittivity of surrounded media. Expressions Eqs.\eqref{shortrange2.2}, corresponding to the $j_x^{I}$ current contribution, still hold with the only difference that now the electron momentum relaxation time Eq.\eqref{Q1} depends on the electron energy
\begin{gather}\label{Сoulomb1}
\frac{1}{\tau_n}=\frac{|n|}{\tau(\varepsilon_{\bf p})},\,\,\,\,
\frac{1}{\tau(\varepsilon_{\bf p})}=\frac{\pi e^4n^C_i}{\hbar\kappa^2\varepsilon_{\bf p}},
\end{gather}
where $n^C_i$ is a concentration of Coulomb centers. The other principal difference of the electron scattering off charged Coulomb centers is the non-zero warping-induced corrections to the collision integral Eq.\eqref{Q3} and the corresponding non-zero contributions to the current density given by $j_x^{II}$ and $j_x^{III}$ in Eq.\eqref{current2}. The direct calculations of these contributions to the current density give
\begin{gather}\label{Сoulomb2}
A^C_{xxx}=
\eta w N\frac{\pi^2 e^5n^C_i}{\kappa^2\hbar}\Bigl[3\langle(\epsilon\tau_{1}^2\tau_{2})'\rangle -\langle(\epsilon^2\tau_{1}^2\tau_{2})''\rangle\Bigr]+\\\nonumber
+\eta w N\frac{\pi^2 e^5n^C_i}{\kappa^2\hbar}\Bigl[\frac{1}{2}\langle(\epsilon\tau_{1}^2\tau_{2})'\rangle+\langle(\epsilon^2\tau_{1}^2\tau_{2})''\rangle\Bigr],
\end{gather}
and
\begin{gather}\label{Сoulomb3}
B^C_{xxx} = \eta w N\frac{\pi^2 e^5n^C_i}{\kappa^2\hbar}\Bigl[3\langle(\epsilon\tau_{1}^2\tau_{2})''\rangle -\langle(\epsilon^2\tau_{1}^2\tau_{2})'''\rangle\Bigr]+\\\nonumber
+\eta w N\frac{\pi^2 e^5n^C_i}{\kappa^2\hbar}\Bigl[\frac{1}{2}\langle(\epsilon\tau_{1}^2\tau_{2})''\rangle+\langle(\epsilon^2\tau_{1}^2\tau_{2})'''\rangle\Bigr].
\end{gather}
Now calculating expressions Eqs.\eqref{shortrange2.2}, accounting for relations Eq.\eqref{Сoulomb1}, and combining the result with contributions \eqref{Сoulomb2}, \eqref{Сoulomb3}, one finally finds
\begin{gather}\label{Сoulomb4}
\chi_{xxx}=\eta\frac{wNk^4\hbar^2}{\pi^2e^7 (n^C_{i})^2}\frac{\langle\epsilon^3\rangle(12-7\pi)}{\rho\left(1-e^{-\frac{N}{\rho T}}\right)}\\\nonumber
\zeta_{xxx} = \eta\frac{wNk^4\hbar^2}{\pi^2e^7 (n^C_{i})^2}\frac{\langle\epsilon^2\rangle(42-21\pi)-(12-7\pi)\frac{\langle\epsilon^3\rangle}{T}e^{-\frac{N}{\rho T}}}{\rho^2\left(1-e^{-\frac{N}{\rho T}}\right)^2}.
\end{gather}

\section{Non-uniform electric field}

Now consider the case when a nonuniform static electric field characterized by the scalar potential $\phi({\bf r})$ is present in the sample. In equilibrium, the electrochemical potential $\mu({\bf r})+e\phi({\bf r})$ of the system must be constant and the current density should vanish. Thus, the current should depend on the spatial derivatives of  $\mu({\bf r})+e\phi({\bf r})$ only. Expanding the current with respect to the derivatives of scalar and chemical potentials and their second-order powers, one obtains
\begin{gather}\label{EF1}
j_i=a_{ij}\nabla_j\mu({\bf r})+b_{ij}\nabla_j\phi({\bf r})+\\\nonumber
+c_{ijk}\nabla_j\nabla_k\mu({\bf r})+d_{ijk}\nabla_j\nabla_k\phi({\bf r})+\\\nonumber
+p_{ijk}\nabla_j\mu({\bf r})\nabla_k\mu({\bf r})
+g_{ijk}\nabla_j\mu({\bf r})\nabla_k\phi({\bf r})+\\\nonumber
+h_{ijk}\nabla_j\phi({\bf r})\nabla_k\phi({\bf r})
\end{gather}
The equality to zero of the total current in equilibrium makes it possible to determine the relation between the coefficients in this expression. These relations replace the Einstein relation in a nonlinear case. Tensors $a_{ij}, b_{ij}$ are known from the linear kinetic phenomena; $c_{ijk}$ and $p_{ijk}$ are derived in the present paper. Non-diagonal components of tensors $a_{ij}, b_{ij}$ determine the VHE transport phenomena and should satisfy the relation of the Einstein type. If the electric field $-\partial_x\phi$ and chemical potential gradient $\partial_xn$ are directed along the $x$-axis, then the VHE coefficients are $a_{xy}=a_{yx}=a_H,\,b_{xy}=b_{yx}=b_H$; and in equilibrium one finds
\begin{gather}\label{EF1.1}
a_H\frac{\partial\mu}{\partial x}+b_H\frac{\partial\phi}{\partial x}=0,
\end{gather}
where $-b_H=\sigma_H$ is the VHE conductivity derived in \cite{VHEGlazovGolub}. Coefficient $a_H$ can be directly found by a comparison with expression Eq.\eqref{Hall7}. That the electro-chemical potential must be constant under equilibrium yields the relation $\partial_x\mu=-e\partial_x\phi$. Combining this statement with Eq.\eqref{EF1.1}, one finds the relation between kinetic coefficients describing the VHE under the electric field and chemical potential gradient as $e\,a_H=-\sigma_H$. The direct comparison of our result Eq.\eqref{Hall7} and $\sigma_H$ expression derived in \cite{VHEGlazovGolub} supports this relation as it must be.

Tensor $h_{ijk}$ components determine the conventional PGE effect in TMD materials \cite{PGEEntin1} and are well known. Thus, the remaining components of $d_{ijk}, g_{ijk}$ tensors can be found from the equilibrium conditions $j_i=0$ and $\mu({\bf r})+e\phi({\bf r})=const$. Taking into account that $\nabla_i\mu({\bf r})=-e\nabla_i\phi({\bf r})$, one finds the relations
\begin{gather}\label{EF2}
eg_{ijk}=e^2p_{ijk}+h_{ijk},\\\nonumber
d_{ijk}=e\,c_{ijk}.
\end{gather}

\section{Conclusion}
To conclude, here we report that the generalized forces given by the electron density gradient (rather than external electric field) lead to the transverse and longitudinal valley linear and nonlinear transport in 2D Dirac monolayer systems preserving the inversion center symmetry. We theoretically demonstrate that the skew electron scattering of diffusive electrons may result in the valley Hall transport as a first order response to the electron density gradient. A theoretical analysis of the second-order system response to the electron density gradients developed here shows the existence of a valley selective PGE-like transport due to the trigonal warping of electron dispersion in the valleys considering the electron scattering processes on both short-range and Coulomb-like impurity centers.

The specific feature of the effects considered here is that the experimental study of these effects does not require two sources of external electromagnetic illuminations as it is for the conventional VHE and PGE transport based on the generalized force given by external electric field. The experimental observation of PGE or VHE transport requires  unequal valley populations by additional circularly-polarized light \cite{VHE2, noneqVHE, PGEEntin1}. In case of the transport effect considered here, the electromagnetic illumination producing the unequal valley populations may simultaneously create the nonuniform electron density distribution, thus, producing the VHE and PGE-like current phenomena studied here.

\section*{Acknowledgement}
This paper was financially supported by the Foundation for the Advancement of Theoretical
Physics and Mathematics “BASIS” and RFBR (project 20-02-00622).

\end{document}